\documentstyle[preprint,aps]{revtex}
\begin{document}


\title{Plasmons in coupled bilayer structures}
\author{S. \ Das Sarma and E. H.\ Hwang}
\address{Department of physics, University of Maryland, College Park,
Maryland  20742-4111 } 
\date{\today}
\maketitle

\begin{abstract}

We calculate the collective charge density excitation dispersion and
spectral weight in bilayer semiconductor
structures {\it including effects of interlayer
tunneling}. The out-of-phase plasmon mode (the ``acoustic''
plasmon) develops a long wavelength
gap in the presence of tunneling with the gap being
proportional to the square root (linear power) of the tunneling
amplitude in the weak (strong) tunneling limit. The in-phase plasmon
mode is qualitatively unaffected by tunneling. 
The predicted plasmon gap should be a useful tool for studying
many-body effects.

\noindent
PACS Number : 73.20.Mf, 73.20.Dx, 71.45.Gm, 71.45.-d

\end{abstract}

\newpage

Collective charge density excitations (or, equivalently plasmon modes)
in bilayer structures have attracted a great deal of theoretical and
experimental attention over the last sixteen years ever since the
existence of an {\it undamped} acoustic (i.e., with a long wavelength
dispersion linear in wave vector) plasmon mode was predicted
\cite{dassarma} in
semiconductor double quantum well systems. In an uncoupled bilayer
system, ignoring any {\it interlayer} Coulomb interaction, each layer
can support a two dimensional (2D) plasma mode \cite{allen} with a
long wavelength 
($q \rightarrow 0$) dispersion $\omega \sim q^{1/2}$ where $q \equiv
|{\bf q}|$, and {\bf q} is the 2D wave vector. When the two layers are
near each other (separated by a distance $d$ in the $z$-direction with
the 2D layers in the {\it x-y} plane), the 2D plasmons are coupled by
the interlayer Coulomb interaction leading to the formation
\cite{dassarma} of
in-phase and out-of-phase interlayer density fluctuation modes
$\omega_{\pm}$: an out-of-phase acoustic plasmon mode, $\omega_-\sim
q$, where the densities in the two layers fluctuate out of phase with
a linear wave vector dispersion, and an in-phase optical plasmon mode,
$\omega_+ \sim q^{1/2}$, where the densities in the two layers
fluctuate in phase with the usual 2D plasma
dispersion. These $\omega_{\pm}$ modes have been observed
\cite{kainth,pinczuk} in double
layer semiconductor quantum well systems via inelastic light
scattering spectroscopic experiments, and the observation of the
$\omega_- \sim q$ mode in GaAs-AlGaAs multilayer 
systems is, in fact, the only unambiguous direct
experimental observation of an acoustic plasmon mode in solid state
plasmas in spite of the theoretical literature on the subject going
back more than forty years \cite{pine}. 

In this Letter we consider the
experimentally relevant issue of the collective mode dispersion in
bilayer structures {\it in the presence of significant interlayer
quantum tunneling}, all earlier work on the subject having dealt with
the zero tunneling limit. It is well-known that tunneling introduces
qualitatively new physics \cite{dassarma1} by introducing a new energy
scale, the {\it interlayer} tunneling energy, in addition to the
Coulomb energy and the {\it intralayer} kinetic energy. We find that
tunneling {\it significantly} affects 
the out-of-phase $\omega_-$ mode, qualitatively modifying its long
wavelength dispersion to $\omega_- \sim (\Delta^2 + C_1 q + C_2
q^2)^{1/2}$ where 
$\Delta$ defines the plasmon gap $\Delta \equiv \omega_-(q=0)$ which
depends nontrivially on the 2D electron density $n$ and the interlayer
tunneling amplitude $t$. In particular we obtain the interesting
result that $\Delta \sim t \;\; {\rm or} \;\; t^{1/2}$ depending on
whether the interlayer quantum tunneling is strong or weak
respectively.  We also find that, in contrast to the no-tunneling
situation when the $\omega_+$ in-phase mode exhausts the plasmon
spectral weight in the long wavelength limit (and the $\omega_-$
acoustic mode carries significant spectral weight only at finite
wave vectors), the out-of-phase mode $\omega_-$ may carry significant
spectral weight in the presence of tunneling even in the long
wavelength limit and may be easily observable via inelastic light
scattering spectroscopy \cite{kainth,pinczuk} or frequency-domain far
infrared (or 
microwave) spectroscopy \cite{allen}. We note the somewhat
non-intuitive result that 
finite tunneling in fact converts the out-of-phase ``acoustic''
plasmon mode (in the $t=0$ situation) to an ``optical'' plasmon mode
(in the $t \neq 0$ situation) by producing a finite plasmon gap
$\omega_-(q=0) = \Delta$ whereas the original ``optical'' plasmon (for
$t=0$) in-phase mode $\omega_+$ becomes the ``acoustic'' plasmon mode,
albeit with a $q^{1/2}$ long wavelength dispersion, in the sense that
$\omega_-(q=0)$ vanishes in the presence of finite tunneling.
The situation in the presence of tunneling is therefore similar to the
familiar phonon terminology where the optical phonon (which has a
finite energy at zero wave vector) corresponds to the out of phase
intracell ionic dynamics and the acoustic phonon (with vanishing long
wavelength energy) corresponds to the in-phase intracell ionic
dynamics. 

The collective mode spectrum is given by the zero of the dynamical
dielectric function of the system, which for a bilayer system in the
presence of a finite interlayer tunneling amplitude $t$ becomes a
tensor $\epsilon_{ijlm}$ of the fourth rank where {\it i, j, l,
m}$=$1 or 2 is the layer index with 1,2 denoting the two layers. The
dielectric function $\epsilon_{ijlm}(q,\omega) \equiv
\delta_{il}\delta_{jm} - v_{ijlm}(q)\Pi_{lm}(q,\omega)$ is obtained
within the mean field random phase approximation (RPA) in our theory
where the deltas are Kronecker delta functions and $v_{ijlm}$ is the
intralayer/interlayer Coulomb interaction matrix element with
$\Pi_{lm}$ as the irreducible noninteracting electron polarizability
function. It is convenient to use one electron energy eigenstates
$E_{\pm}$ (we take $\hbar =1$ throughout this paper) as the basis set
rather than the layer index since the latter is {\it not} a good
quantum number in the presence of tunneling. The energy levels
$E_{\pm} = \varepsilon(k) \pm t$, where $\varepsilon(k) = k^2/2m$ is
the parabolic one electron 2D kinetic energy in each layer and $t$ is the
tunneling strength, are  the usual symmetric and
antisymmetric states in the presence of tunneling with a single
particle symmetric-antisymmetric (SAS) gap given by $\Delta_{SAS} =
E_+ - E_- = 2t$. In the SAS representation the collective mode spectra
become decoupled by virtue of the symmetric nature of our double
quantum well system (i.e., both layers identical with equal electron
density), and the collective density fluctuation spectra are given by
the following two equations for the in-phase and the out-phase plasmon
modes $\omega_{\pm}$ respectively:
\begin{equation}
\epsilon_{+}(q,\omega) = 1-v_+(q) \left [ \Pi_{++}(q,\omega) +
\Pi_{--}(q,\omega) \right ] = 0,
\label{ep}
\end{equation}
and
\begin{equation}
\epsilon_{-}(q,\omega) = 1-v_-(q) \left [ \Pi_{+-}(q,\omega) +
\Pi_{-+}(q,\omega) \right ] = 0.
\label{em}
\end{equation}
In Eqs. (\ref{ep}) and (\ref{em}) the Coulomb interaction matrix
elements are given by $v_{\pm}(q) = v_1(q) \pm v_2(q)$ where
$v_{1,2}(q)$ are respectively the intralayer and interlayer Coulomb
interaction matrix elements. We use the simplest model for the Coulomb
interaction (at no loss of generality) assuming the intralayer Coulomb
interaction to be purely a 2D Coulomb interaction (and thus neglecting
subband effects in each layer, which is entirely justified in most
experimental situations where the intralayer intersubband energy is much
larger than $\Delta_{SAS}$) -- subband effects can be trivially
incorporated by using the appropriate subband form factors
\cite{jain}. For this 
simple model, $v_1(q) = 2\pi e^2/(\kappa q)$; $v_2(q) = v_1(q)
\exp(-qd)$, with $\kappa$ as the (high frequency) background lattice
dielectric constant. Finally, $\Pi_{\alpha \beta}(q,\omega)$ in
Eqs. (\ref{ep}) and (\ref{em}), with $(\alpha,\beta)=(+,-)$, are the
noninteracting SAS polarizability functions within our RPA theory:
\begin{equation}
\Pi_{\alpha \beta}(q,\omega) = 2\int \frac{d^2k}{(2\pi)^2}
\frac{f_{\alpha}({\bf k}+{\bf q}) - f_{\beta}({\bf k})}{w +
E_{\alpha}({\bf k}+{\bf q}) - E_{\beta}({\bf k})},
\label{piab}
\end{equation}
where $f_{\alpha,\beta}$ are Fermi occupancy factors (we restrict
ourselves to $T=0K$ in this paper), and the factor of two in the front
arises from spin.

Solving Eqs. (\ref{ep})--(\ref{piab}) we obtain the collective density
fluctuation spectra of the coupled bilayer system. In the absence of
tunneling, $t=0$, one has $E_+=E_-=\varepsilon({\bf k})$, and one then
recovers in a straightforward fashion the well-known optical
($\omega_+ \sim q^{1/2}$) and acoustic ($\omega_- \sim q$) plasmons of
a bilayer system without any electron tunneling. It is, in fact,
straightforward (but quite tedious) to obtain analytically [from
Eqs. (\ref{ep})--(\ref{piab})] the long
wavelength ($q \rightarrow 0$) plasma modes of the coupled bilayer
system {\it including effects of interlayer tunneling}. We obtain in the
long wavelength limit the following results:
\begin{equation}
\omega_{+}^2(q \rightarrow 0) = \frac{2\pi e^2 N}{\kappa m} q,
\label{wp}
\end{equation}
\begin{equation}
\omega_-^2(q \rightarrow 0) = \Delta^2 + C_1 q + C_2 q^2,
\label{wm}
\end{equation}
where $N=2n$ is the total 2D electron
density ($n$ being the electron density per layer), $C_2 > 0$, and
\begin{eqnarray}
\Delta^2 & = &\Delta_{SAS}^2 + \frac{\pi}{m} (n_+-n_-) (q_{TF}d)
\Delta_{SAS} \nonumber \\
C_1 & = & -(\Delta^2 - \Delta_{SAS}^2)\frac{d}{2}
\label{del}
\end{eqnarray} 
where $q_{TF} = 2me^2/\kappa$ is the 2D Thomas-Fermi wave vector and
$n_{\pm}=n \pm n_c$ is the electron density in the
symmetric/antisymmetric $\pm$
level. Here, $n_c = (m/2\pi) \Delta_{SAS}$ for $n > n_c$ when both
symmetric and antisymmetric levels are occupied (i.e., the 2D Fermi
energy $E_F > \Delta_{SAS}$), and $n_c = n$ for 
$n \leq  n_c$ when only the symmetric level is  occupied (i.e., $E_F <
\Delta_{SAS}$).  
Note that the in-phase mode $\omega_+$ is unaffected in the long
wavelength limit either by finite tunneling or by level occupancy, and
depends only on the total 2D electron density $N$, following the
standard 2D plasmon dispersion. The positive coefficient
$C_2(q_{TF},d,E_F,\Delta_{SAS})$ is not shown for brevity.

The most important qualitative feature of the plasmon dispersion in
the presence of interlayer tunneling 
is that the out-of-phase mode $\omega_-$, which
is purely acoustic in the absence of tunneling ($\omega_- \sim q$ if
$\Delta_{SAS} = 0$), develops a plasmon gap at $q=0$ in the presence
of nonzero tunneling. It is easy to see from Eqs. (\ref{wm}) and
(\ref{del}) that this
plasmon gap $\Delta$,  $\omega_-(q=0) = \Delta$,
has the following behavior
\begin{equation}
\Delta \sim \Delta_{SAS} \;\; {\rm or} \;\; \sqrt{\Delta_{SAS}},
\end{equation}
depending on whether the interlayer tunneling is strong
($\Delta_{SAS}/E_F \gg q_{TF}d$) or weak ($\Delta_{SAS}/E_F \ll
q_{TF}d$). It 
should be emphasized that the strikingly non-intuitive $\Delta_{SAS}^{1/2}$
dependence of the collective mode gap (on the square root of the
single particle gap) is purely a Coulomb interaction effect, which
dominates the collective excitation spectra in the weak tunneling
situation. Finally, it is interesting to note that the first order
dispersion correction to the out-of-phase plasmon gap is negative.

In Figs. \ref{fig1} -- \ref{fig3} we present our numerical results for
the collective excitation spectra of the coupled bilayer system
without restricting to the long wavelength limit. In our calculations
we have used RPA and also the so-called Hubbard approximation (HA)
which includes a model static local field correction \cite{jonson} to the
noninteracting RPA irreducible polarizability [Eq. (\ref{piab})]. In
general, HA should be a better approximation than RPA at lower
electron densities (and higher wave vectors) although, being an
uncontrolled approximation, the quantitative improvement in HA over
RPA is unknown. In Figs. \ref{fig1} and \ref{fig2} we show our
calculated collective 
mode dispersions for different electron densities (with a fixed
interlayer separation of $d=200\AA$) in both RPA and HA, and for both
zero and non-zero tunneling ($\Delta_{SAS}=0$, $\Delta_{SAS}=1$
meV). The important features to note in these results are: (1) In
general both tunneling and local field correction have little effect
on the in-phase mode $\omega_+$, which even at low densities seems to
be well described by the long wavelength RPA formula
[Eq.(\ref{wp})]. This is a direct consequence of the $f$-sum rule --
the $\omega_+$ mode being the effective 2D plasma mode of the system
is robust and insensitive to many body and/or
tunneling effects. This somewhat disappointing result is however
important because it states emphatically that all efforts to study
many-body effects by studying the usual in-phase 2D plasma dispersion are
doomed to failure, a fact already empirically known to
experimentalists. (2) The out-of-phase $\omega_-$ mode is 
strongly influenced by tunneling and local field effects, and could in
principle be a sensitive experimental tool in studying many-body
effects \cite{hu}. (3) Local field effects in general reduce the $\omega_-$
frequency, and at low densities this could even lead to a complete
suppression of the $\omega_-$ mode. 
In the absence of tunneling this complete suppression of $\omega_-$
mode occurs below the critical density $n_c =
q_{TF}^2/[8\pi(1+q_{TF}d)^2]$ in the HA. 
Tunneling, however, has the dramatic effect of making the suppressed
$\omega_-$ mode reappear when the layer separation $d$ increases above
$d_c = 1/2m\Delta_{SAS} - 1/q_{TF}$  even if $n<n_c$. 
(Note that, within the  RPA, this Landau damping induced suppression
of $\omega_-$ mode 
does not happen, and the $\omega_-$ mode exists for small wave vectors
at all densities.) 
(4) Tunneling opens up a zero frequency plasmon gap in the
$\omega_-$ mode. In RPA the calculated gap $\Delta \equiv
\omega_-(q=0)$ is always 
greater than $\Delta_{SAS}$ [Eq. (\ref{del})]. 
In the HA, however, $\Delta$ could be above or below $\Delta_{SAS}$,
and is given by
\begin{equation}
\Delta^2  = \Delta_{SAS}^2 + \frac{\pi}{m} (n_+-n_-)( q_{TF}d) \left (
1 - \frac{1}{2k_Fd} \right ) \Delta_{SAS}.
\end{equation}
If the total density $N < N_c = 1/(8\pi d^2)$, the plasmon gap
$\Delta$ is less than the single particle gap $\Delta_{SAS}$, and we
have the interesting situation of a collective plasmon mode being
lower in energy than the corresponding single particle energy. 
(5) Tunneling leads to a weak negative dispersion of the $\omega_-$ mode
at long wavelengths (i.e., $C_1 < 0$). (6) Finally, a very interesting finding
(Fig. \ref{fig2}) is that at low densities the $\omega_-$ mode is, in
fact, more stable in the HA (than in RPA) because it lies below the
single particle 
continuum in the presence of tunneling. In the absence of
tunneling the $\omega_-$ mode may reappear at {\it large wave vectors}
beyond the single particle continuum although it is completely
suppressed (by Landau damping) at long wavelengths. This reappearance
of the $\omega_-$ 
mode at large wave vectors (for $\Delta_{SAS}=0$) is purely a local
field effect, and has earlier been attributed \cite{neilson} to a
charge density wave 
instability in bilayer structures. Tunneling, according to our theory,
stabilizes the long wavelength $\omega_-$ mode which now lies between
the two (symmetric and antisymmetric) single particle continua, and is
in fact more stable in HA than in RPA. Tunneling, therefore, opposes
the presumptive charge density wave instability \cite{neilson}.

In Fig. \ref{fig3} we show our calculated spectral
weight or the dynamical structure factor $S(q,q_z,\omega)$, which
is given \cite{jain} by the imaginary part of the density-density
correlation function where $q_z$ is the probe wave vector normal to
the layers \cite{kainth,pinczuk,jain}, for the collective modes in
coupled bilayer systems both in RPA and HA for finite
tunneling. For $q_z d=0.0$ only  the in-phase mode $\omega_+$ carries
any weight,
but for finite $q_z d$  the out-of-phase mode carries substantial
spectral weight even at long wavelengths ($q \rightarrow 0$).
The most important message here is that both $\omega_{\pm}$
modes in general carry finite spectral weights and should be observable
in resonant inelastic light scattering \cite{kainth,pinczuk} and far
infrared optical \cite{allen} spectroscopies. 

We point out that the $\omega_-$ mode, which presumably becomes the
Goldstone mode in the symmetry-broken phase, could be used
as an experimental probe to search for the theoretically predicted
\cite{zheng} interlayer-spontaneous-phase-coherent quantum state in
low density bilayer structures.
It is interesting in this context to point out that it has been
claimed \cite{kalman} that the local field correction by itself within
the so-called quasilocalized charge approximation \cite{kalman}, {\it
even in the absence of any interlayer tunneling}, could lead to the
opening of a long wavelength gap in the out of phase mode in bilayer
systems.

We note that the main approximations of our theory, namely treating
the tunneling amplitude $t$ as a phenomenological parameter and using
the Hubbard approximation for the local field correction, should not
in any way affect our qualitative conclusions. In particular, $t$ 
decreases with increasing layer separation $d$ in a calculable
way, and therefore for large $d$ one always recovers the no-tunneling
limit of a vanishing plasma gap. While there is no general consensus
on the best possible local field corrections in electron liquids, it
is well-known \cite{jonson} that the Hubbard approximation gives
quantitatively similar results to more sophisticated local field
corrections involving self-consistent approximations\cite{jonson}. 
We emphasize that the existence of a plasma
gap in the presence of tunneling and its asymptotic dependence on $t$
are independent of our approximations.
 
In conclusion, we establish in this paper that bilayer semiconductor
double quantum well structures should be a useful tool for
studying the interplay between tunneling and many-body effects on
collective mode dispersion -- in particular, the dispersion of the
out-of-phase collective mode in the presence of tunneling should show
interesting observable many-body effects.

This work is supported by the U.S.-ARO and the U.S.-ONR. One of us
(S.D.S) thanks J. Eisenstein for asking an important question.

\begin{figure}
\caption{RPA plasmon dispersion for (a) $n=10^{11} {\rm
cm}^{-2}$, and (b)  $n=10^{10} {\rm cm}^{-2}$. Solid
(dashed) lines represent plasmon dispersion in the absence (presence)
of  interlayer tunneling. Inset shows that the finite wave vector
minimum of the out-of-phase mode $\omega_-$ in the presence of tunneling.
We use parameters  
corresponding to GaAs quantum wells: $m= 0.067 m_e$, 
$\kappa = 10.9$, and $d = 200 \AA$, $\Delta_{SAS} = 1.0$ meV for all
our figures. Shaded region indicates the single particle excitation
(SPE) Landau damping continuum.} 
\label{fig1}
\end{figure}

\begin{figure}
\caption{Local field effects on the plasmon dispersion for density $n
= 1.0 \times 10^9 {\rm cm}^{-2}$: (a) in the
absence of tunneling, (b) in the presence of tunneling ($\Delta_{SAS} =
1.0$ meV). Inset shows the reappearance of the out-of-phase mode
$\omega_-$ near $q=2k_F$. }
\label{fig2}
\end{figure}

\begin{figure}
\caption{The dynamical structure factor for (a) $n=10^{10} {\rm
cm}^{-2}$ and (b) $n=10^{9} {\rm cm}^{-2}$ in the RPA (solid
lines) and HA (dashed lines) for finite 
tunneling ($\Delta_{SAS}=1.0$ meV). Here, thin (thick) lines correspond to 
$q_z d = 0.0$ ($q_z d = \pi/4$). }
\label{fig3}
\end{figure}

\end{document}